# Black phosphorus as a new broadband saturable absorber for infrared passively Q-switched fiber lasers

*Tian Jiang[1,2,3*], Ke Yin[1], Xin Zheng[1], Hao Yu[1], Xiang-Ai Cheng[1,2]*

[1] College of Optoelectronic Science and Engineering, National University of Defense Technology, Changsha 410073, China
[2] State key laboratory of High Performance Computing, National University of Defense Technology, Changsha 410073, China
[3] State Key Laboratory of Low-Dimensional Quantum Physics, Department of Physics , Tsinghua University, Beijing
10084, P. R. China
[*] *jiangtian198611@163.com*

Black phosphorus (BP) with its enticing electric and optical properties is intensely researched in the field of optoelectronics. In this paper, Q-switched pulses at 1550 nm and 2 μm wavelengths are obtained by inserting bulk-structured BP based saturable absorber (SA) into an erbium-doped fiber laser (EDFL) and an thulium/holmium-doped fiber laser (THDFL), respectively. The BP-SA was prepared by depositing powered BP material on to the flat side of a side-polished single mode fiber. Q-switched 1550 nm pulses with width tuned from 9.35 to 31 μs were obtained for the EDFL. For the THDFL, over 100 nm wavelength range could be achieved from 1832 to 1935 nm by adjusting the pump power. To the best of our knowledge, these results demonstrated the broadband saturable absorption property of BP and for the first time verified that BP as a new two-dimensional material for applications in saturable absorption devices.

**OCIS** *codes*: (060.4370) Nonlinear optics, fibers; (160.4330) Nonlinear optical materials; (140.3540) Lasers, Q-switched.

**1. Introduction**

Current research in new SA primarily focuses on layered structure materials [1-16]. The advancement of grapheme [1, 2] in SA motivates the exploration of other layered structure materials , such as topological insulators (TIs) [3-11, 17] and transition metal dichalcogenides (TMDCs) [12-14, 16]. TIs are a new class of materials that have a bulk band gap and gapless Dirac surface/edge state [18], which is protected by the topological symmetry. Since Bernad et al. [15] first discovered the saturable absorption of $Bi_2Te_3$ TI in 2012, several experiments demonstrated the utilization of different TI SAs for passive mode-locking and Q-switching at 1, 1.55 and 2 μm over a very broadband spectral region. Very recently, the saturable absorption of molybdenum disulphide ($MoS_2$) [14], as a member of TMDCs [12-14], was also demonstrated using the Z-scan technique at 800 nm. The $MoS_2$ in monolayer form becomes a direct band gap (1.8 eV) semiconductor in the visible range, whereas its gap (0.86-1.29 eV) is indirect for the multilayered material [12]. By inserting into the $MoS_2$ SA into erbium-doped fiber laser (EDFL), a ~710 fs pulse centered at 1.569 μm wavelength with a repetition rate of 12.09 MHz was demonstrated [16]. However, most of the TMDCs currently being studied have band gap large than 1 eV [19], rendering them unsuitable for mid-infrared applications.

Black phosphorus (BP) is the most stable phosphorus allotrope [19-21]. Similar to graphene, BP is a layered material in which individual atomic layers are stacked together by van der Waals interactions [22, 23]. The band gap of BP increases with the decreasing number of layers from 0.3 eV in bulk state to 2 eV in monolayer BP [19, 20]. This characteristic is similar to $MoS_2$, except that in BP the band gap is always direct for all number of layers. Therefore, the narrow gap of multi-layer BPs can fulfill the space between grapheme's zero gap and TMDCs' large gap, making it an ideal layered structure material for near and mid-infrared optoelectronics [19-21]. However, recent studies of BP mostly focus on electronics

[24], there are rare reports caring about their optical applications. Very recently, Hanlon et al. dispersed few-layered BP nano-sheets in amide solvents and demonstrated their broadband nonlinear absorption ability at 1030 nm and 515 nm using the open-aperture Z-scan technique. Their results show that BP exhibits much stronger SA response than graphene at both wavelengths. Also inspired by the success of $MoS_2$ in nonlinear fiber lasers, Lu et al have tried their best efforts to demonstrate the very broadband SA of multi-layer BP ranging from the visible (400 nm) towards mid-infrared region (at least 1930 nm). With the demonstrations of BP's SA characteristics, it is foreseeable and prospective that they could also be used in nonlinear optics as other layered structure materials.

In this paper, BP material is confirmed to have saturable absorption response in the near-infrared region experimentally. The BP based SA was prepared by depositing bulk-structured BP particles onto a side-polished single mode fiber with the presence of a scotch tape. The BP was then demonstrated to be a broadband optical intensity modulator (Q-switcher) in both an erbium-doped fiber laser (EDFL) at the 1550 nm and a thulium/holmium-doped fiber laser (THDFL) around 2 μm. By inserting the BP-SA into the EDFL, Q-switched pulses with maximum pulse energy of 28.3 nJ is achieved, the shortest pulse width is 9.35 μs. When the BP-SA is inserted into a THDFL, Q-switched pulses with shortest pulse width of 2.53 μs and highest output pulse energy of 276 nJ at 1890 nm are obtained. To the best of our knowledge, this is the first report on BP-SA based passively Q-switched fiber lasers and the successful demonstration of BP material's broadband saturable absorption in real fiber laser implementations.

## 2. Preparation and characteristics of BP-SA

The original crystal BP is provided by University of Science and Technology of China, and it has the similar dimension of a rice grain. The crystal BP is then mechanically polished into many micrometers scale powders with a commercial abrasive paper (the surface roughness of the abrasive paper is 9 μm). Then, a scotch tape is used to transfer these powders and deposit them onto the flat side of a side-polished single mode fiber to make the BP-SA. Comparing to the other well-known methods used in preparing 2D materials like mechanical exfoliation, ultrasound-assisted liquid-phase exfoliation and chemical vapor deposition, the great advantages of this method are of low cost, great repeatability and easy of manipulation.

Figure 1(a) shows the morphology of the BP powders characterized by using scanning electron microscopy (SEM). It is clear that these powders are small particles with sizes of less than 10 μm and a little irregular. Then it could be inferred that these BP powders own the bulk-structure characteristics rather than layered-structure as prepared by liquid exfoliation method reported in [25, 26]. The measured Raman spectrum of the bulk-structured BP is depicted in Fig. 1(b). There are three Raman optical photon peaks at wave numbers of 361, 437 and 465.5 $cm^{-1}$, which is in good accordance with previous reports [21, 27]. Figure 1(c) shows the optical microscope images of the surface of the scotch tape before it is deposited onto the side polished fiber. The red dashed line in Fig. 1(c) divides the areas with and without BP materials. Figure 1(d) shows the microscope image of the flat side of the side-polished single mode fiber. It is obvious that the measured width of the flat side is ~120 μm, and the distance between the flat side and the fiber core is calculated to be ~13 μm. The measured insertion loss of the side-polished fiber before the deposition of BP materials is ~0.5 dB at 1550 nm, which originates from the leakage of evanescent waves at the polished region. This leakage ensures the efficient interaction between the intra-cavity light filed and the BP materials not only at the wavelength of 1550 nm but also the longer 2 μm region, since that the fundamental mode diameter expands with the increasing of laser wavelength.

## 3. Laser results and discussion

### 3.1 Passively Q-switched EDFL at 1550 nm

Firstly, the aforementioned BP-SA was incorporated into a ring-cavity EDFL. Figure 2 shows the experimental setup of the Q-switched EDFL. The all fiber oscillator consists of a 2 m erbium-doped fiber (EDF), an optical intensity isolator (ISO), a polarization controller (PC), the BP-SA, a 20:80 optical coupler (OC) and a 980/1550 nm wavelength division multiplexer (WDM). The EDF has a core/cladding diameter of 3/125 μm with a core numerical aperture (NA) of 0.28. A single mode laser diode (LD) with maximum output power of ~500 mW at 976 nm is used to pump the EDF via the WDM. A fiber pigtailed band pass filter (BPF) at 1550 nm is used to stabilize the operating wavelength. Unidirectional light propagation in the ring oscillator is ensured by the optical polarization independent ISO. All these rest passive fibers are standard single mode fibers (SMF, Corning SMF-28). It has to note that the side-polished fiber based BP-SA is polarization-dependent [8, 28], so that the PC is employed to adjust the coupling of light from the fiber to the absorber. Output of the EDFL is extracted from the 20% port of the OC. The measuring equipment includes an optical spectrum analyzer (Yokogawa, AQ6375), an optical power meter (AV6334), a fast InGaAs photo-detector (Thorlabs, DET10D/M) together with a digital oscilloscope (Tektronix, TDS7154).

The laser started to emit Q-switched pulses at the pump power of 10.4 mW. With careful adjustment of the PC, the signal voltages on the oscilloscope could be maximized. Stable Q-switched laser pulses were observed with the pump power up to 24.3 mW. Figure 3 shows typical output characteristics of the Q-switched EDFL. Figure 3(a) plots the measured pulse train with pump power of 11.7, 15.5 and 21.2 mW. It is observed that when the pump power is 17 mW, the output pulse width is 21.8 μs and the output pulse repetition rate is 7.06 kHz. With the increasing of pump power, the output pulses become narrowing and the pulse repetition rate increases. The output spectrum of the Q-switched EDFL is shown in Fig. 3(b) measured at a resolution of 0.05 nm, which locates at 1550 nm with a 3 dB spectral bandwidth of about 0.15 nm, and the optical signal to noise ratio is better than 50 dB, indicating that the laser performance enabled by the BP-SA is stable.

Next, the output pulse characteristics of the Q-switched EDFL under different pump power were investigated in details. Figure 4(a) shows the evolutions of output average power and pulse peak power with the increasing of pump power. In the experiment, when the pump power got higher than 25 mW, the Q-switched pulses disappeared and the output turned to be continuous wave operation. But when the pump power was decreased, the Q-switched pulses reappeared. This phenomenon may relate to the physical characteristics of BP, but the exact reason is not clear to us now. The output pulse width and pulse repetition rate as functions of the pump power are plotted in Fig. 4(b). It is shown that typical features of Q-switched fiber lasers are observed in the BP-SA based EDFL. The output pulse repetition rate increased almost linearly with the increase of the pump power. The repetition rate can be adjusted in the range of 4.43 to 18 kHz. The pulse width is as long as 31 μs at 4.43 kHz, and it decreases to 9.35 μs when the pump power gets to 24.3 mW.

In the experiment, we also checked the role of the BP-SA in the Q-switched EDFL by removing it. This method has been widely used to make sure whether the SA responses for Q-switched or mode-locked operation of fiber lasers [4, 7]. Even though the pump power was adjusted and the PC was rotated in a large range, only continuous wave operation could be observed after removing the BP-SA. This demonstrates that our BP-SA responses for the effective Q-switched operation of the EDFL.

*3.2 Wavelength-tunable Q-switched THDFL at 2 μm region*

In order to test the broadband saturable absorption of BP, the as-mentioned BP-SA was also incorporated into a THDFL which performs at the 2 μm wavelength region. The detailed experimental setup is depicted in Fig. 5 which is similar to the 1550 nm laser setup as shown in Fig. 2. The gain medium is a 2 m length of single mode thulium/ holmium-doped fiber (THDF, Coractive) pumped by a home-made 1550 nm continuous-wave fiber laser via a 1550/2000 nm WDM. The THDF has a core/cladding diameter of 9/125 μm, and a core NA of 0.18. Both the ISO and OC operate at the 2 μm wavelength region. Different from the previous Q-switched EDFL, here a tunable filter is used to filter and stabilize the lasing wavelength. The grating based tunable filter is designed to work at the 2 μm wavelength region. It has a 3 dB spectral bandwidth of ~1.7 nm. The output power is measured by a thermal power meter (Thorlabs, S302C).

With the increasing of the pump power, Q-switched pulses at 2 μm wavelength region could be obtained. Fig. 6(a) shows the output pulse trains of Q-switched THDFL at the operating wavelengths of 1860, 1890 and 1920 nm under the same pump power of 730 mW. It is shown that the repetition rate is as high as 73 kHz at the working wavelength of 1860 nm, and then decreases to 45 kHz at 1890 nm, and finally decreases to only 35.3 kHz at 1920 nm. Figure 6(b) plots the corresponding output spectra of the Q-switched THDFL centered at 1860, 1890 and 1920 nm, respectively. The corresponding 3 dB bandwidths of the generated Q-switched pulses are measured to be 0.13, 0.12 and 0.18 nm.

Further, the performance of the Q-switched THDFL was investigated by adjusting the pump power and meanwhile the tunable filter. Table 1 includes the detailed characteristics of the Q-switched THDFL, together with the results of Q-switched EDFL. Benefited from the tunable filter, the output wavelength of the Q-switched THDFL could be continuously tuned over 100 nm, ranging from 1832 to 1935 nm. However, due to the comparatively large insertion loss of the tunable filter and the BP-SA, the wavelength tunability of the Q-switched THDFL was limited to only ~100 nm, but enough to verify BP's broadband SA characteristics at the 2 μm region. As provided in Table. 1, the obtained shortest pulse width and maximum pulse energy were 2.53 μs and 276 nJ, respectively.

From Table 1, it is shown that the tuning ranges of the output pulse width and repetition rate at 1832 and 1935 nm are much narrower than at other three wavelengths. The limited tuning ranges of the Q-switched THDFL mainly arise from the gain provided by the THDF rather than the broadband saturable absorption ability of BP-SA. With sufficient gain afforded by the THDF and essential methods to minimize the as-mentioned loss, it is believed that the operating wavelength range of the Q-switched THDFL could be further extended.

*3.3 Discussion*

Firstly, in order to make sure that the Q-switchings of both the EDFL and the THDFL were indeed benefited from the BP-SA, contrast tests were conducted. At this time, only clean scotch tape was placed onto the side-polished fiber, whatever the pump power was adjusted and the PC position was rotated in the experiments, only continuous wave laser were observed. The results demonstrate that BP materials are responsible for the Q-switched pulse generation in the two fiber lasers at 1550 nm and the 2 μm region. Secondly, due to the limited experimental condition at present there are no direct measurement of the saturable absorption properties of the as-mentioned BP-SA, such as the information of its linear absorption ability and saturable modulation depth, but it is also enough and sufficient to demonstrate BP's efficient SA characteristics by applying it to Q-switched fiber lasers directly. Further, the broadband SA characteristics were demonstrated from 1550 nm to ~2000 nm in this paper by combining the experimental results of the EDFL and the THDFL utilizing one side-polished fiber based BP-SA. Thirdly, despite being the most stable allotrope of phosphorus, BP materials are very hydrophilic and will disappear in ambient conditions with oxygen and water molecules. This might leads to the degradation of BP's performance. Fortunately, like other 2D layered materials BP-SA could be isolated from the environment and be protected by mixing them with other infrared transparent organic materials for long-term stable performance, such as polymethyl methacrylate and polyvinyl acetate.

As the criterion of stable mode locking operation of a laser above a certain Q-switched threshold shows [29],

$$E_p^2 > E_{sat,g} E_{sat,a} \Delta R \qquad (1)$$

where the $E_p$ represents the intra-cavity pulse energy, $E_{sat,g}$ is the saturation energy of the gain fiber, $E_{sat,a}$ is the saturation energy of the SA, and $\Delta R$ represents the modulation depth of the SA. However, BP has efficient optical absorption ability as reported in [25, 26] resulting in high modulation depths of few-layered BP which were measured to be as high as 19.5% and 16.1 % at the wavelength of 1563 nm and 1930 nn, respectively. Due to the bulk-structured nature of the BP-SA used in this paper, so that the modulation depth $\Delta R$ of the BP-SA could be inferred to be much higher when compared to few-layered BP materials. The high value of $\Delta R$ might eventually change the direction of the inequality, so that no mode-locking pulses were observed during the experiments. It is expected that with few-layered or monolayer BP materials [30], ultrafast mode-locking operation of fiber lasers at the infrared region could be realized, which will be our future work.

## 4. Conclusion

BP material has been demonstrated as a broadband SA in the near-infrared wavelength region by incorporating a bulk-structured BP-SA into two different fiber lasers. With the BP-SA, Q-switched laser pulses were obtained in both the EDFL at 1550 nm and the THDFL at the 2 μm wavelength region. The ability to fabricate SA, combined with the fact of BP's broadband saturable absorption in the infrared region, makes it a candidate for future optoelectronic applications. It is also discussed that due to BP's stronger optical absorption ability only few-layered or single-layered BP materials might be used for mode-locking operation of fiber lasers, but bulk-structured BP materials are more suitable for Q-modulator or light switcher applications.

**Acknowledgement**

The authors thank Associate Professor Zhangqi Song for his kind support of side-polished single mode fibers. This work is supported by the Chinese National Science Foundation (NO: 61340017).

**Figures and tables**

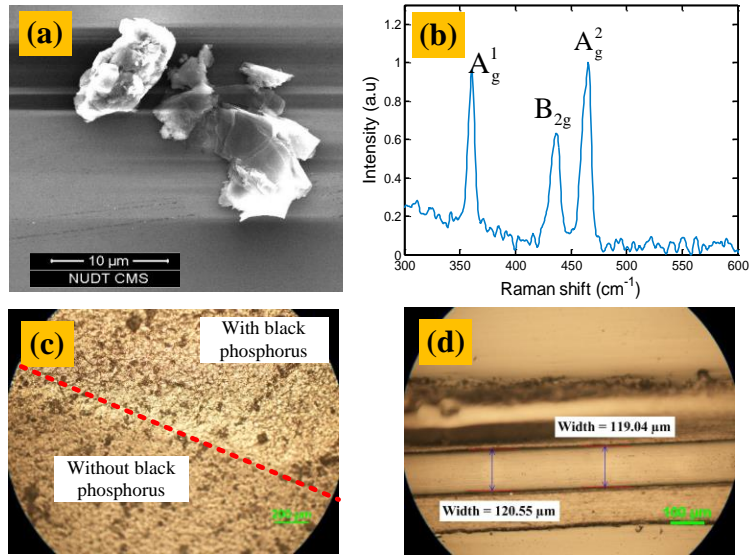

Fig.1 (a) SEM photograph of the bulk-structured BP powders. (b) Raman spectrum of bulk-structured BP. (c) Optical microscope image of the BP powders on the scotch tape. (d) Optical microscope image of the flat side of the side-polished single mode fiber.

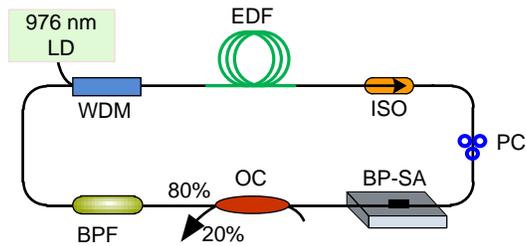

Fig.2 The experimental setup of the passively Q-switched EDFL.

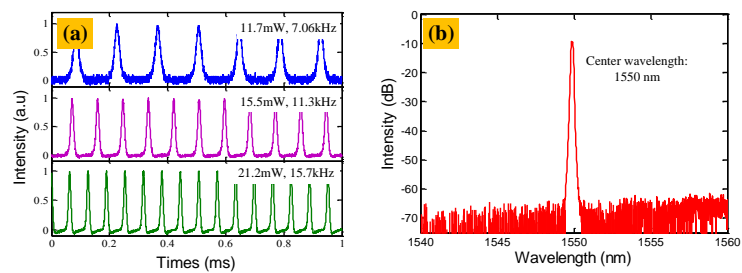

Fig. 3 Output characteristics of the Q-switched EDFL. (a) Pulse train. (b) Optical spectrum.

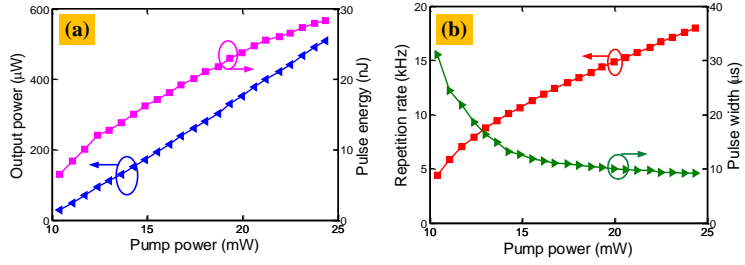

Fig. 4 (a) Output average power and pulse energy as functions of pump power. (b) Pulse repetition rate and duration versus incident pump power.

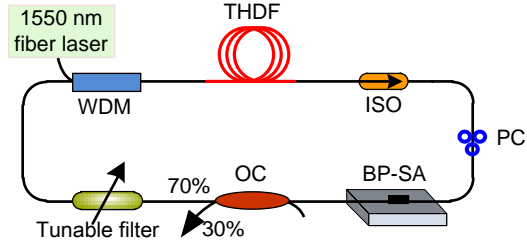

Fig.5 The experimental setup of the passively Q-switched THDFL.

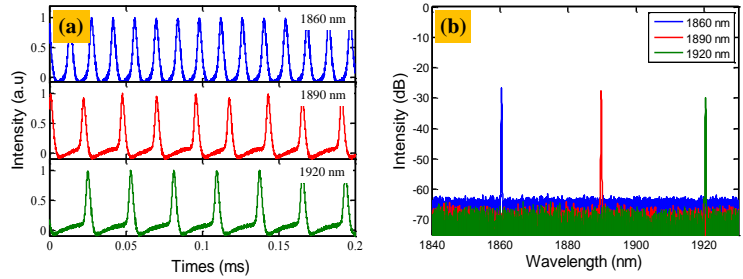

Fig.6 Output characteristics of Q-switched THDFL. (a) Pulse train. (b) Optical spectrum.

Table 1 Characteristics of the BP-SA based Q-switched fiber lasers.

| Laser setup | Wavelength (nm) | Repetition rate (kHz) | Pulse width (μs) | Maximum pulse energy (nJ) |
|---|---|---|---|---|
| Q-switched EDFL | 1550 | 4.43-18 | 9.35-31 | 28.3 |
| Q-switched THDFL | 1832 | 20-25.5 | 4-6.67 | 75 |
|  | 1860 | 26.4-73 | 3.1-4.2 | 80.8 |
|  | 1890 | 24-57.2 | 2.53-4.8 | 276 |
|  | 1920 | 14-60 | 3-4.7 | 238 |
|  | 1935 | 20-42 | 4.9-5.7 | 114 |